\documentclass[11pt]{article}
\usepackage{graphicx} 
\usepackage{amsmath}
\usepackage{amsfonts}
\usepackage{amssymb}
\usepackage[T1]{fontenc}
\usepackage[utf8]{inputenc} 
\usepackage{subcaption}
\usepackage{multirow}
\usepackage{url}
\usepackage{lmodern}
\usepackage{textcomp}
\usepackage{bm}
\usepackage{booktabs} 
\usepackage{caption}
\usepackage{float}
\newcommand{\mat}[1]{\mbox{\boldmath{$#1$}}}
\usepackage{natbib}
\usepackage{color}
\usepackage{hyperref} % Gerenciamento de hiperlinks
\title{Diagnosing overdispersion in longitudinal analyses with grouped nominal polytomous data}
\author{Maria Letícia Salvador$^{1}$ \\
    Gabriel Rodrigues Palma$^{2, 3}$ \\
    Rafael de Andrade Moral$^{2, 3}$ \\
    Idemauro Antonio Rodrigues de Lara$^{1}$ }
\date{}

\begin{document}

\maketitle

\noindent{} 1. Luiz de Queiroz College, University of S\~{a}o Paulo, Piracicaba, Brazil;

\noindent{} 2. Hamilton Institute, Maynooth University, Maynooth, Ireland;

\noindent{} 3. Department of Mathematics and Statistics, Maynooth University, Maynooth, Ireland;
\vspace{1cm}

\begin{footnotesize}
    %\SingleSpacing
    \noindent \small{\textbf{Abstract:}}
    \noindent \small 
    
Experiments in Agricultural Sciences often involve the analysis of longitudinal nominal polytomous variables, both in individual and grouped structures. Marginal and mixed-effects models are two common approaches. The distributional assumptions induce specific mean-variance relationships, however, in many instances, the observed variability is greater than assumed by the model. This characterizes overdispersion, whose identification is crucial for choosing an appropriate modeling framework to make inferences reliable. We propose an initial exploration of constructing a longitudinal multinomial dispersion index as a descriptive and diagnostic tool. This index is calculated as the ratio between the observed and assumed variances. The performance of this index was evaluated through a simulation study, employing statistical techniques to assess its initial performance in different scenarios. We identified that as the index approaches one, it is more likely that this corresponds to a high degree of overdispersion. Conversely, values closer to zero indicate a low degree of overdispersion. As a case study, we present an application in animal science, in which the behaviour of pigs~(grouped in stalls) is evaluated, considering three response categories.
%palavras-chave
    
 
 \end{footnotesize} 

\section{Introduction}

Nominal polytomous variables are those defined by a finite set of two or more distinct categories. These variables are of interest in various fields, especially in Agricultural Sciences, where experiments are often designed with experimental units such as animals, to quantify their behaviour or plants, to classify them according to their morphology. In certain situations, the experimental unit may consist of a fixed group of individuals, such as a stall containing multiple animals. For such cases, categorical data with grouped structures are frequently considered.

According to \cite{agresti2019} the most common distribution associated with polytomous nominal data is the multinomial distribution, which belongs to the multiparametric exponential family. Models involving the multinomial distribution are multivariate extensions of generalized linear models \citep{nelder1972}. Additionally, there might be interest in exploring repeated measures or observations on the same individuals over a certain period, characterizing a longitudinal study. This approach allows for a more detailed analysis of the evolution of the response variable over time, allowing for the incorporation of different types of correlation structures between observations \citep{molenberghs2005}.

The most common models for longitudinal data analysis are marginal, transition, and mixed models \citep{diggle2002, fitzmaurice2008longitudinal, hand2017practical}. Mixed generalized linear models \citep{breslow1993, salinas2023generalized}, combining mixed-effects models and generalized linear models \cite{zeger1988models, stiratelli1984random}. These models assume that the correlation between individual responses results from individual heterogeneity. Additionally, the model allows the identification of correlations within individuals, and the inclusion of fixed and random effects. 

Specifically for a polytomous nominal response, with individual or grouped data, some extensions of mixed models, using generalized logits are presented by
\cite{hartzel2001multinomial, chan2023multilevel}. In particular, with grouped data, the inclusion of pertinent random effects in the linear predictor structure can explain the heterogeneity between individuals in groups and the serial structure of observations. \citep{weijs2020analysing, liu2021older}.

%{\color{red}To improve accuracy and robustness, the estimation of fixed effects and variance components can be done using Restricted Maximum Likelihood (REML) \citep{chan2023multilevel}. This approach allows for a more straightforward interpretation of within-class dynamics, which is important for hierarchical data \citep{mcfadden1000mixed} está solto esse parágrafo}.

A fundamental aspect of these models is that the observed variance is expected to be close to the variance assumed by the model. However, this requirement is not always achieved. When the observed variance is greater than the expected by the model, this phenomenon is referred to as overdispersion \citep{hinde1998overdispersion, demetrio2014models}. Recognizing the presence of overdispersion in the data is crucial for adopting appropriate measures and selecting a model that ensures more precise estimation. Ignoring the issue of overdispersion can lead to inadequate model fit, and parameter standard error estimates are underestimated, compromising model inferences \cite{olsson2002generalized}.

For count data, \cite{ridout1998models} introduced a dispersion index that evaluates the presence of overdispersion, calculated as the ratio between the mean and the variance. When this index equals one, the data are considered equidispersed; however, if it is greater than one, the presence of overdispersion in the data is suggested. Based on this premise, \cite{salvador2022analysis} proposed a dispersion index that detects this phenomenon in cross-sectional studies for grouped nominal polytomous data. This index compares the observed variance of the data with the model's expected variance.

Thus, this work aims to extend \cite{salvador2022analysis}'s work and initial exploration for the construction of a longitudinal multinomial dispersion index, aiming to provide a robust descriptive and diagnostic tool for detecting overdispersion in grouped nominal polytomous data. The relevance of this index lies in its ability to enhance the interpretation of models that may be affected by this dispersion characteristic. The initial phase of this work will be dedicated to conducting simulation studies, and employing statistical techniques to evaluate the initial performance of this index in different scenarios. This work is structured as follows: a review of the multinomial distribution and mixed-effects generalized linear models for nominal polytomous data in the longitudinal context, and the causes of the overdispersion phenomenon, are presented in Section \ref{methods}. Subsequently, Section \ref{IDML} introduces the novel longitudinal multinomial dispersion index proposal. We present simulation studies to evaluate its effectiveness and the results obtained in Section \ref{simulation}. We illustrate the methodology using a real data application in Section \ref{application}. Finally, we present concluding remarks in Section \ref{conclusion}.

%Thus, this work aims to extend \cite{salvador2022analysis}'s work and propose a longitudinal multinomial dispersion index, which acts as a descriptive and diagnostic measure of the presence of the overdispersion phenomenon in nominal polytomous data with a clustered structure. This work is structured as follows: a review of the multinomial distribution and mixed-effects generalized linear models for nominal polytomous data in the longitudinal context, and the causes of the overdispersion phenomenon, are presented in Section \ref{methods}. Subsequently, Section \ref{IDML} introduces the novel longitudinal multinomial dispersion index proposal. We present simulation studies to evaluate its effectiveness and the results obtained in Section \ref{simulation}. We illustrate the methodology using a real data application in Section \ref{application}. Finally, we present concluding remarks in Section \ref{conclusion}.

\section{Methods} \label{methods}

\subsection{Review of Classical Methods for Polytomous Data}

\subsubsection{Multinomial Distribution}

%According to \cite{agresti2019}, the multinomial distribution is most commonly associated with experimental scenarios where the random variable is categorical and polytomous, and it is more frequently used in the class of generalized linear models with a polytomous response variable.

Consider a random experiment in which the sample space admits $J$ possible outcomes, mutually exclusive, $A_1, A_2, \cdots, A_J$. Therefore, each realization of the experiment will result in only one of the events with probability $\pi_j=P(A_j)$, where $\displaystyle \sum_{j=1}^{J} \pi_{j}=1$. Let $n$ be the number of realizations of this trial and further let the response vector be denoted by $\mat{Y}$, where $\mat{Y} = (Y_1, Y_2, \cdots, Y_J)^{'}$, where each component $Y_j$ describes the number of times, that the events $A_j$ occur such that $\displaystyle \sum_{j=1}^{J} y_{j}=n$.

Thus, the distribution of $\mat{Y}$ is multinomial and its probability function is given by:

\begin{equation} \label{distmult}
P[Y_1=y_1, \cdots, Y_J=y_J] = f(\mat{y}\mid\mat{\pi})= \dfrac{n!}{\displaystyle \prod^{J}_{j=1}y_{j}!}\displaystyle \prod^{J}_{j=1} \pi_j^{y_j}
\end{equation}

\noindent where $y_j \in {0, 1, \cdots, n}$, $\mat{\pi} = (\pi_1, \pi_2, \cdots, \pi_J)^{'}$ with $0 \leq \pi_j \leq 1$.

From the multinomial distribution (\ref{distmult}), for each category $j$ the result $y_j$ has mean and variance given by $\mbox{E}(Y_j)=n\pi_j$ and $\mbox{Var}(Y_j)=n\pi_j(1-\pi_j)$, respectively. However, the covariance between $y_j$ and $y_k$, $\forall j \neq k$, $j=1, \cdots, J$ and $k=1, \cdots, J$ is given by $\mbox{Cov}(Y_j, Y_k)= -n\pi_j \pi_k$, note that the covariance is negative due to the constraint $\displaystyle \sum_{j=1}^{J} y_{j}=n$.

\subsubsection{Mixed Generalized Logits Model} \label{sec:cap2_mixedlogit}

To establish notation, let $\mat{Y}{ijt}$ be the vector of response variables for the $i$-th individual in the $j$-th response category at time $t$ ($i=1,2,\cdots, n$, $j=1,2,\cdots, J$, $t=1,2,\cdots, T$). Also consider that $\mat{x}{it} = (\mat{x}_1, \cdots, \mat{x}p)^{'}$ is a vector of $p$ covariates corresponding to individual $i$ at time $t$. The probability function of the response variable vector $\mat{Y}{ijt}$ conditional on the vector $\mat{u}_i$, belongs to the multivariate exponential family with a linear predictor that includes the random vector $\mat{u}_i = (u_1, \cdots, u_n)^{'}$ at the individual level, where $\mat{u}i$ follows a multivariate normal distribution $ \mat{\mbox{N}}_{n} (\mat{0}, \mat{\Sigma}_j)$, with $\mat{\Sigma}_j$ being the variance-covariance matrix of dimension $n \times n$

Let $\mbox{Y}_{ijt}$ be the vector of polytomous response variables, with $j=1, \cdots, J-1$ categories and $J-1$ logits, where $\mbox{Y}_{ijt}$ follows a multinomial distribution, i.e., $\mat{Y}_{ijt} \sim \mbox{Multinomial} (\mat{y}_{ijt}, \mat{\pi}_{ijt})$. Setting the $J$-th response category as the reference, the model is defined by \citep{hedeker2003mixed}:

\begin{equation} \label{MLGM}
    \ln \left[\dfrac{\pi_{ijt}}{\pi_{iJt}}\right] = \eta_{ijt}= \mat{x^{'}_{it}} \mat{\beta_j} + \mat{z^{'}_{it}} \mat{u_i}
\end{equation}

\noindent where $\eta_{ijt}$ is the linear predictor of the $j$-th logit, $\mat{x_{it}}$ is the covariate vector associated with the fixed effects parameters $\mat{\beta_j} = (\beta_{1_j}, \beta_{2_j}, \cdots, \beta_{p_j})$, and $\mat{z_{it}}$ represents the terms associated with the random effects $\mat{u}_i$. Furthermore, the random effects are assumed to follow a normal distribution, $\mat{u}_i \sim \mat{\mbox{N}}_{n} (\mat{0}, \mat{\Sigma}_j)$, with dimension $n \times n$.

According to \cite{chan2023multilevel}, the penalized log-likelihood function, $l=l_1+l_2$, conditional on the random effect $\mat{u_i}$, is given by:

\[l_1 = \sum_{i=1}^{n}\sum_{t=1}^{T}\sum_{j=1}^{J-1} \delta _{ijt} \eta_{ijt} - \sum_{i=1}^{n}\sum_{t=1}^{T}\sum_{j=1}^{J-1} \log \left[1 +  \sum_{j=1}^{J-1} \exp(\eta_{ijt})\right] \]

\noindent where $\delta _{ijt} = 1$ if $\mbox{Y}_{ijt}=j$ and $\delta _{ijt} = 0$ otherwise. For the $i$-th individual, in the $j$-th category at time $t$, $\sum_{j=1}^{J}  \delta _{ijt} \equiv 1, \forall i, j, t$.

The penalty function $l_2$, which is the logarithm of the conditional probability function on $\mat{u_i}$, is defined by:
\[l_2 = - \dfrac{1}{2} \left[  n\log(2 \pi \sigma^2) + \dfrac{1}{2} \mat{u}^{'} \mat{u} \right] \]

Considering the model (\ref{MLGM}), the probabilities predicted by the model are defined by:

\begin{equation} \label{prob_MLGM}
   \displaystyle{\hat{\pi}}_{ijt} = 
\begin{cases}
\dfrac{\exp(\displaystyle{\hat{\eta}}_{ijt})}{1 +  \sum_{j=1}^{J-1} \exp(\displaystyle{\hat{\eta}}_{ijt})}, \hspace{0.2cm} \mbox{if} \hspace{0.1cm} j=1,2,\cdots, J-1 \\
 \dfrac{1}{1 +  \sum_{j=1}^{J-1} \exp(\displaystyle{\hat{\eta}}_{ijt})}, \hspace{0.2cm} \mbox{if}, \hspace{0.1cm} j=J
\end{cases} 
\end{equation}

\noindent where $\hat{\eta_{ijt}}=  \ln \left[\frac{\hat{\pi}_{ijt}}{\hat{\pi}_{iJt}}\right] $, with $j=1,2,\cdots, J-1$.

To diagnose overdispersion, tools such as the residual deviance and dispersion indices adapted to different types of data, such as counts \citep{ridout1998models} and grouped nominal polytomous data in cross-sectional contexts \citep{salvador2022analysis}, are applied. However, these methodologies have limitations, especially when dealing with longitudinal data. We now present a new statistic to diagnose overdispersion based on the mixed generalized logits model.

\subsection{Longitudinal Multinomial Dispersion Index} \label{IDML}

In this section, we describe the construction of a dispersion index, aiming to obtain a diagnostic measure to assess the degree of overdispersion present in nominal polytomous data grouped in longitudinal studies. To establish notation, let us first define grouped data. Let $Y_{ijt}$ be the response variable vector for the $i$-th sampling unit, each comprising a group of individuals of fixed size $m$ ($i=1, \cdots, n$, $j=1, \cdots, J$, $t=1, \cdots, T$). Therefore, $Y_{ijt}$ quantifies the occurrences of category $j$ within each group $m_i$ at time $t$, such that the sum of occurrences across all categories in the group equals the total number of individuals in the group, i.e., $\sum^{J}_{j=1} Y_{ijt} = m$. Additionally, $Y_{ijt} = (Y_{i1t}, Y_{i2t}, \cdots, Y_{iJt})^{'}$ follows the multinomial distribution, with parameters $m$ and $\mat{\pi}_i = (\pi_{i1t}, \pi_{i2t}, \cdots, \pi_{iJt})^{'}$. The structure of grouped nominal polytomous data in the longitudinal context is presented in Table \ref{tab_ind}.

\begin{table}[H]
\centering
\footnotesize
\caption{Structure of nominal polytomous grouped data in longitudinal studies, illustrating the occurrence of $j$ categories in each experimental unit ($i$), where $m_i$ indicates the number of individuals per group, observed over $t$ distinct periods.}
\begin{tabular}{cccc}
\hline 
Experimental unit ($i$) & $m$ & Response Vector & Covariate Vector \\
\hline 
1 & $m$ & $\mat{y}_{1t} = (y_{11t}, y_{12t}, \cdots, y_{1Jt})^{'}$ & $\mat{x}_{1t} = (x_{1t1}, x_{1t2}, \cdots, x_{1tp})^{'}$ \\
2 & $m$ & $\mat{y}_{2t} = (y_{21t}, y_{22t}, \cdots, y_{2Jt})^{'}$ & $\mat{x}_{2t} = (x_{2t1}, x_{2t2}, \cdots, x_{2tp})^{'}$ \\
$\vdots$ & $\vdots$ & $\vdots$ & $\vdots$ \\
$n$ & $m$ & $\mat{y}_{nt} = (y_{n1t}, y_{n2t}, \cdots, y_{nJt})^{'}$ & $\mat{x}_{nt} = (x_{nt1}, x_{nt2}, \cdots, x_{ntp})^{'}$ \\
\hline
\end{tabular}
\label{tab_ind}
\end{table}

The proposed dispersion index is based on the ratio between the observed data variance and the variance assumed by the model, in this case, the mixed generalized logits model. The observed data variance is given by:

\begin{equation} \label{VO}
    \mat{\mbox{V}}^{\mbox{o}}_{jt} = \dfrac{1}{n-1} \displaystyle \sum_{i=1}^{n} (\mat{y}_{ijt}-\Bar{y}_{ijt}).
\end{equation}

Considering the mixed generalized logits model defined in (\ref{MLGM}), the expected variance by the model is expressed by:

\begin{equation} \label{VE}
    \mat{\mbox{V}}^{\mbox{e}}_{jt} = n \hat{\pi}_{jt} (1- \hat{\pi}_{jt}),
\end{equation}

\noindent where $\hat{\mat{\pi}}_{jt}$ was defined in equation (\ref{prob_MLGM}).

Based on these definitions, the longitudinal multinomial dispersion index is constructed as follows: 

\begin{enumerate} 

\item[1.] Initially, a dispersion index is calculated for each category $j$ with respect to each time occasion $t$, that is, 

\begin{equation} \label{ins_passo1}
\mat{\Lambda}_{jt} = \frac{\mbox{V}^{\mbox{o}}_{jt}}{\mbox{V}^{\mbox{e}}_{jt}}
\end{equation}

\noindent where $\mbox{V}^{\mbox{o}}_{jt}$ is the observed variance in the data for the $j$-th category at time $t$, and $\mbox{V}^{\mbox{e}}_{jt}$ is the expected variance by the model for the $j$-th category at time $t$, defined in equations (\ref{VO}) and (\ref{VE}), respectively.

\item[2.] The mean of the dispersion indices obtained in (\ref{ins_passo1}) is calculated for each response category, resulting in the dispersion index for each time occasion:

\begin{equation} \label{ind_passo2}
\mat{\Lambda}t = \dfrac{\sum_{j=1}^{J}\mat{\Lambda}_{jt}}{\mbox{J}}
\end{equation}

\noindent where $t=1, \cdots, T$ and J is the total number of response categories.

\item[3.] The mean of the dispersion indices obtained in (\ref{ind_passo2}) is calculated with respect to the time occasions,

\begin{equation} \label{ind_passo3} 
\mat{\Lambda}_{m} = \dfrac{\sum{t=1}^{T}\mat{\Lambda}_{t}}{T}, \end{equation}

\noindent where $T$ is the total number of time occurrences of the response categories.

\item[4.] Finally, the dispersion index $\mat{\Lambda}_m$ obtained in (\ref{ind_passo3}) is divided by the number of individuals in group $m$, so that the longitudinal multinomial dispersion index is defined by:

\begin{equation}
\Lambda_{\mbox{\footnotesize longitudinal}} = \dfrac{\mat{\Lambda}_{m}}{m}, \end{equation}

\noindent where $m$ is the number of individuals in the group.

\end{enumerate}

\subsection{Simulation Study} \label{simulation}

To assess the effectiveness of the proposed longitudinal dispersion index for nominal polytomous data with grouped structure, a simulation study was conducted.

We considered $54$ scenarios considered, given by combinations of sample sizes $N\in\{100, 200,500\}$, number of response categories ($j \in \{3, 4, 5\}$), number of individuals per group ($m \in \{5, 10, 15\}$), and number of repeated measurements over time $t \in \{3,4\}$. For each scenario, $1000$ datasets were simulated. Furthermore, simulations were conducted based on mixed generalized logits models, where the structure of the linear predictor consists of an effect of a continuous covariate, in addition to the intercept, including a random effect that captures the dependency of individuals over time.

For the model, the sample variables were simulated from:

\begin{equation} \label{mod_indice}
    \ln \left[\dfrac{\pi_{ijt}}{\pi_{i1t}}\right] = \alpha_{jt} + \beta_j x_{it} + z_{it} u_i, \hspace{0.2cm} j=2, \cdots, J
\end{equation}

\noindent where $x_{it}$ are realizations of a standard normal random variable with $t=3$ and $4$ time occasions, and $u_i$ represents the intra-individual random effect, with $u_i \sim N(0, \sigma^{2})$. 

%Through the random effect $\mathbf{u}$, we can induce scenarios of equidispersion and overdispersion, by fixing $\sigma^{2}=0.01$ or $\sigma^{2}=10$, respectively.

The parameter values were defined as follows:

\[\mat{\theta}_{(J=3)} = (\alpha_{2}, \alpha_{3}, \beta_{2}, \beta_{3}) = (1,0; 0,5; 0,5; 1,0),\]

\vspace{-0.5cm}

\[\mat{\theta}_{(J=4)} = (\alpha_{2}, \alpha_{3}, \alpha_{4}, \beta_{2}, \beta_{3}, \beta_{4}) = (1,0; 0,5; 1,5; 0,5; 1,0; -1,0) \hspace{0.2cm} \text{and}\]

\vspace{-0.5cm}

\[\mat{\theta}_{(J=5)} = (\alpha_{2}, \alpha_{3}, \alpha_{4}, \alpha_{5}, \beta_{2}, \beta_{3}, \beta_{4}, \beta_{5}) = (1,0; 0,5; 1,5; 1,0; 0,5; 1,0; -1,0; -0,7)\]

In this simulation study, we investigated the ability of the longitudinal multinomial dispersion index to diagnose overdispersion in longitudinally grouped nominal polytomous data. Equidispersion and overdispersion scenarios were introduced by fixing $\sigma^{2}=0.01$ or $\sigma^{2}=10$, respectively, through the random effect $\mat{u}$ in model (\ref{mod_indice}). Simulations were carried out using R software \citep{R}, employing the \texttt{mclogit} package \citep{elff2022package}, suitable for fitting mixed-effects generalized logits models. The index performance was assessed by calculating statistical measures such as mean and standard deviation. Additionally, we verified whether the simulated indices across all scenarios exhibited symmetric behaviour and if there were modifications in the considered scenarios, using the Shapiro-Wilk normality test \citep{shapiro1965analysis}. The calculation method of these indices is detailed in Section \ref{IDML}. All code and simulated data are provided at\url{https://github.com/GabrielRPalma/PolytomousDataOverdispersionIndex}.

\subsection{Simulation Study Results}

Initially, we compare the results of the longitudinal multinomial dispersion indices for the cases of equidispersion ($\sigma^{2} = 0.01$) and overdispersion ($\sigma^{2} = 10$), based on $1000$ simulations conducted for a sample size of $N=100$, in accordance with Tables \ref{IDML_TabM5} to \ref{IDML_TabM15}.

\begin{table}[H]
\centering
\footnotesize
\caption{Descriptive statistics related to the longitudinal dispersion index based on $1000$ sets of simulated data, with groups composed of $5$ individuals ($m=5$), for sample size $N=100$.}
\begin{tabular}{lrr|rr|rr}
\hline
& \multicolumn{2}{c}{$j=3$} & \multicolumn{2}{c}{$j=4$} & \multicolumn{2}{c}{$j=5$} \\ \cline{2-7} 
 & \multicolumn{6}{c}{$t=3$}  \\ \cline{2-7} 
 & $\sigma^{2} = 0.01$   & $\sigma^{2} =10$   & $\sigma^{2} = 0.01$   & $\sigma^{2} =10$   & $\sigma^{2} = 0.01$   & $\sigma^{2} =10$  \\ 
 \hline
Maximum   & 0.375 & 0.720  &  0.340 & 0.640 &  0.320  & 0.625 \\
Minimum   & 0.280 & 0.613  & 0.242 & 0.533 & 0,240  & 0.515 \\
Amplitude  & 0.094 & 0.107  & 0.098 & 0.108 & 0.079  & 0.110 \\
Mean  & 0.328 & 0.668  & 0.287 & 0.589 & 0.282  & 0.570 \\
Standard deviation & 0.016 & 0.017  & 0.013 & 0.016 & 0.013  & 0.017 \\
\hline
 & \multicolumn{6}{c}{$t=4$} \\ \cline{2-7} 
 & $\sigma^{2} = 0.01$   & $\sigma^{2} =10$   & $\sigma^{2} = 0.01$   & $\sigma^{2} =10$   & $\sigma^{2} = 0.01$   & $\sigma^{2} =10$  \\ 
 \hline
Maximum     & 0.373 & 0.724  & 0.332 & 0.643 & 0.323  & 0.620 \\
Minimum  & 0.284 &  0.622 & 0.251 & 0.538 & 0.240  & 0.524 \\
Amplitude & 0.088 & 0.102  & 0.081 & 0.104 & 0.082  & 0.095 \\
Mean      & 0.327 & 0.669  & 0.287 & 0.590 & 0.282  & 0.571 \\
Standard deviation & 0.014 & 0.015  & 0.011 & 0.014 & 0.011  & 0.014 \\
\hline
\end{tabular}
\label{IDML_TabM5}
\end{table}

\begin{figure}[H]
\centering
\includegraphics[width=1.0\textwidth]{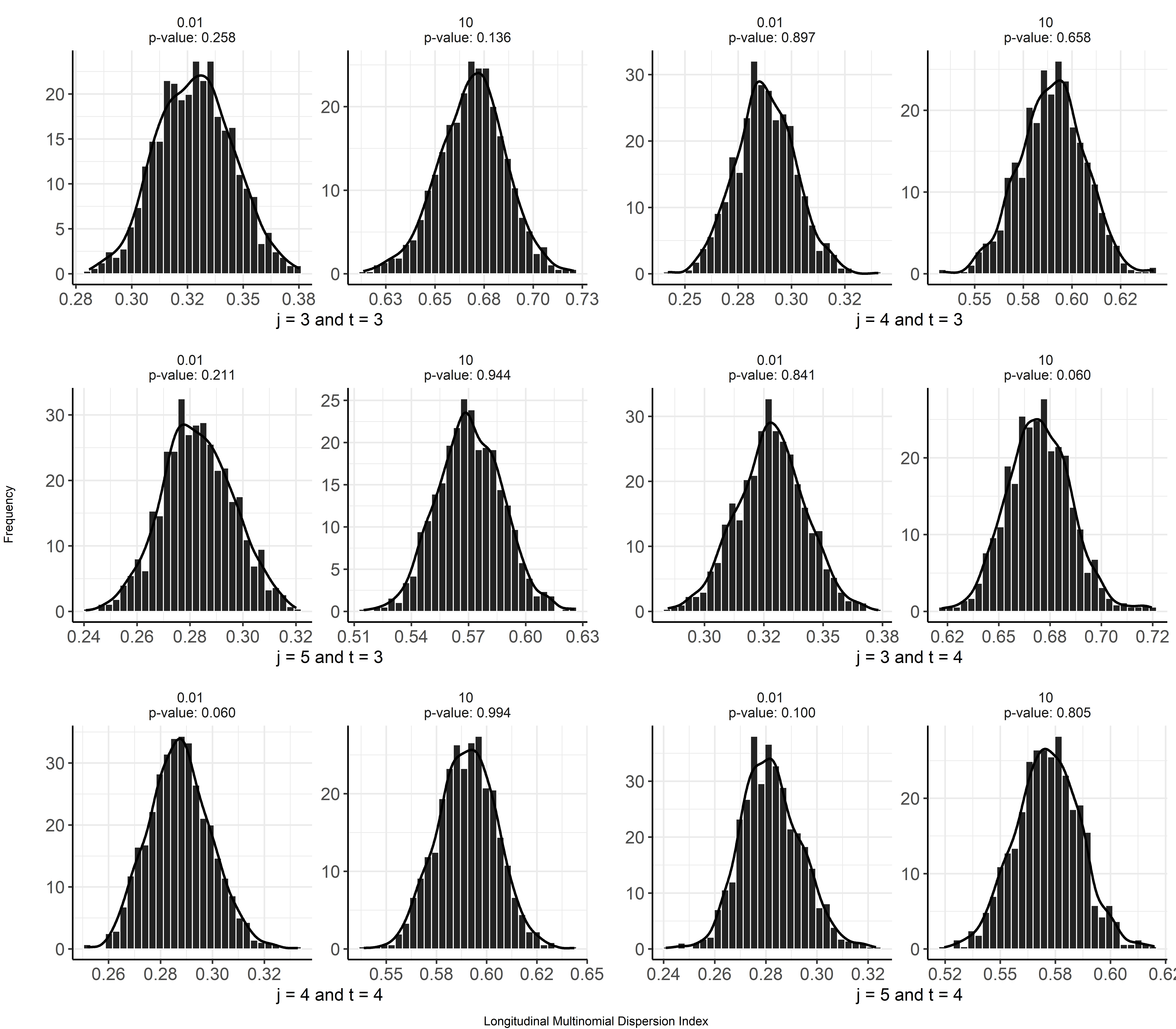}
 \caption{Histogram of the longitudinal dispersion index based on $1000$ simulated data sets, with groups consisting of 5 individuals ($m=5$), for a sample size of $N=100$.}
 \label{IDML_HistM5}
\end{figure}

\begin{figure}[H]
\centering
\includegraphics[width=1.0\textwidth]{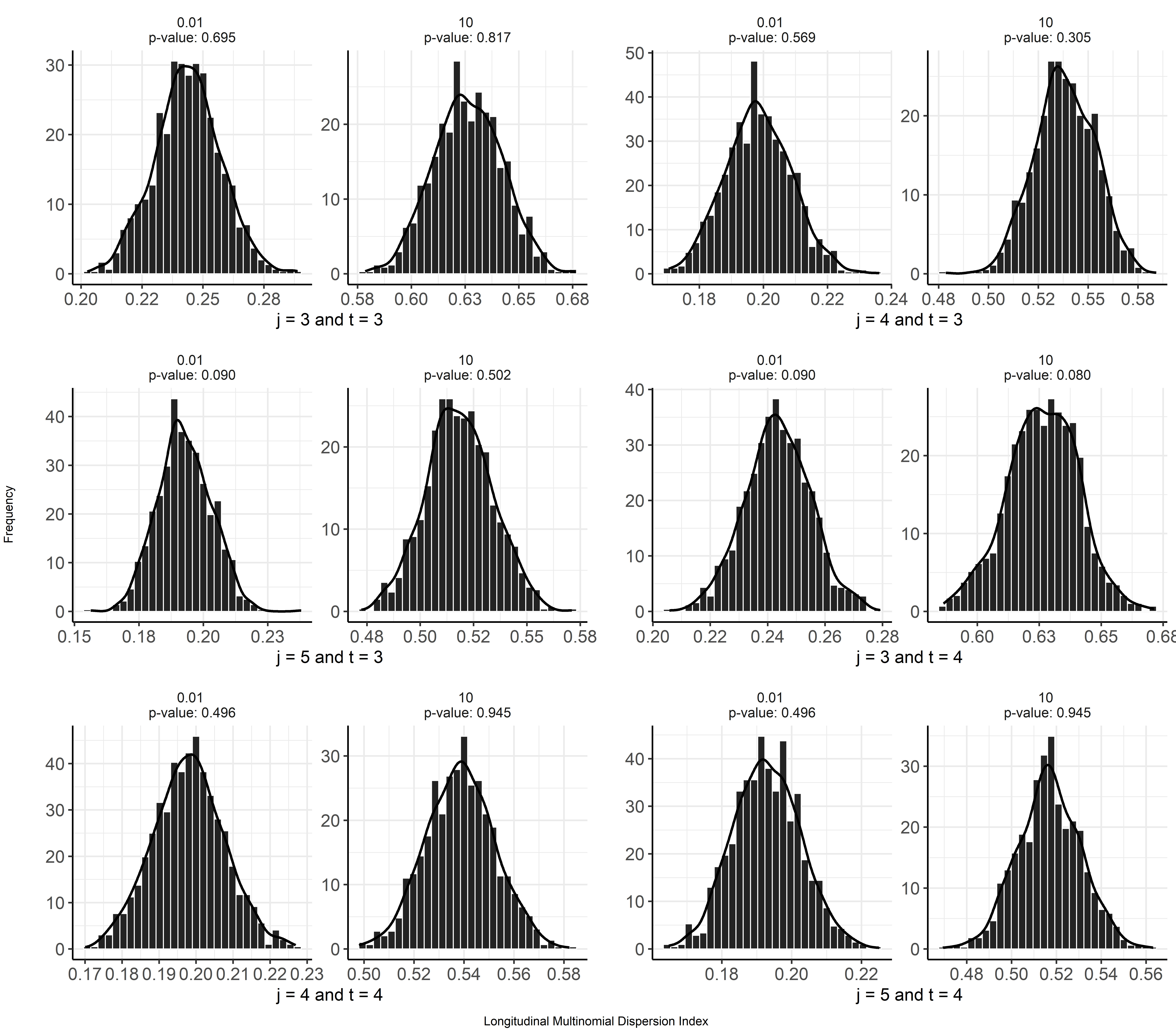}
 \caption{Histogram of the longitudinal dispersion index based on $1000$ simulated data sets, with groups consisting of 10 individuals ($m=10$), for a sample size of $N=100$.}
 \label{IDML_HistM10}
\end{figure}

\begin{figure}[H]
\centering
\includegraphics[width=0.9\textwidth]{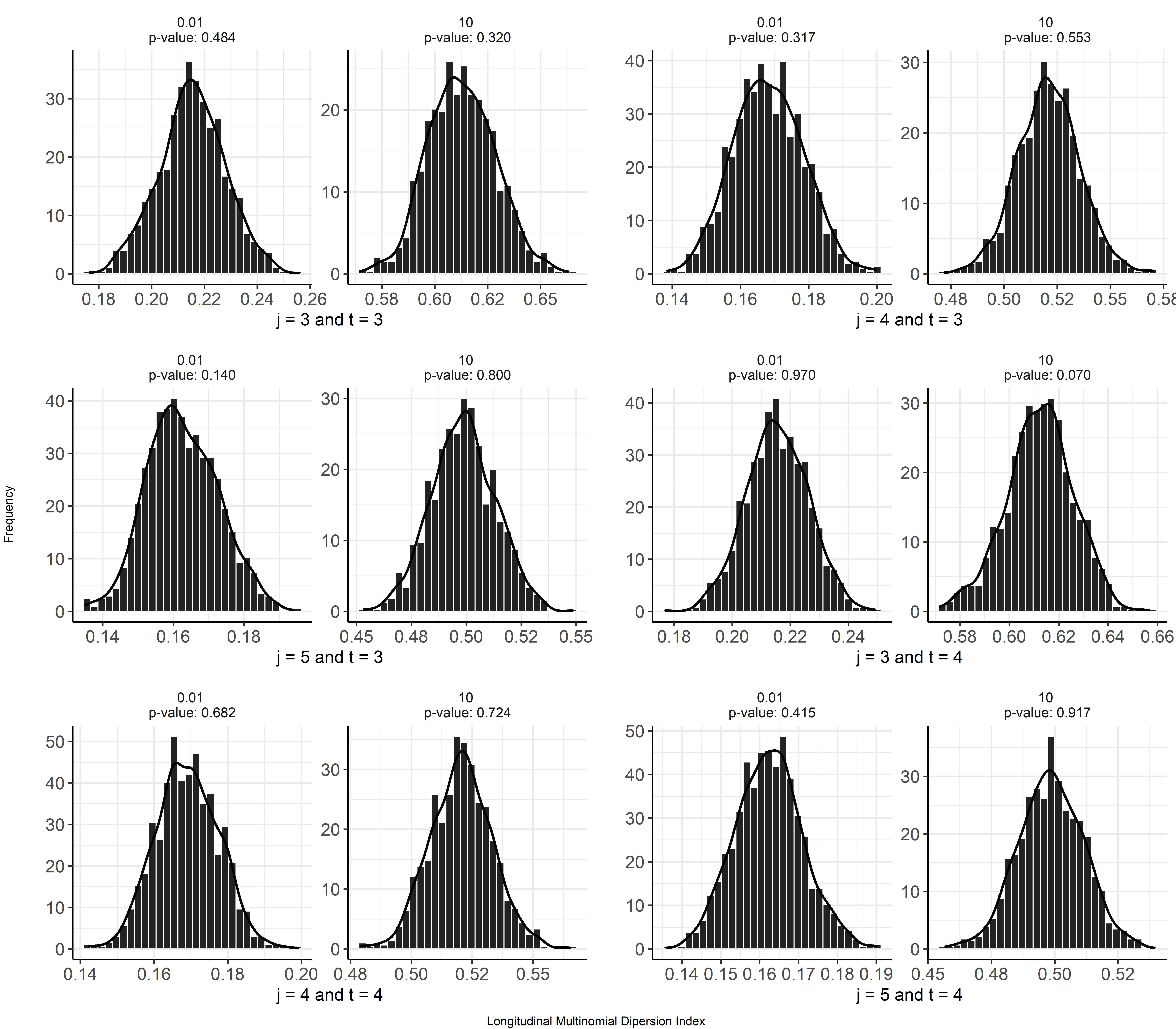}
 \caption{Histogram of the longitudinal dispersion index based on $1000$ simulated data sets, with groups consisting of 15 individuals ($m=15$), for a sample size of $N=100$.}
 \label{IDML_HistM15}
\end{figure}

Initially, scenarios involving groups composed of $m=5$ individuals were considered. The results, as presented in Table \ref{IDML_TabM5}, show that, for example, for $j=3$ response categories and $t=3$ time points, the dispersion index ($\mat{\Lambda}_{\mbox{\footnotesize longitudinal}}$) varied approximately from $0.280$ to $0.375$ in scenarios considered to be equidispersed ($\sigma^{2} = 0.01$) and from $0.613$ to $0.720$ in scenarios with degrees of overdispersion ($\sigma^{2} =10$). Furthermore, when comparing the simulations performed for $\sigma^{2} = 0.01$ and $\sigma^{2} = 10$, it was found that the average dispersion index in all $\sigma^{2} = 10$ scenarios was approximately twice as high as that in the scenarios with $\sigma^{2} = 0.01$, indicating that the higher values clearly provide evidence of overdispersion in these cases. The histograms, illustrated in Figure \ref{IDML_HistM5}, demonstrate that the distribution of the dispersion index ($\mat{\Lambda}_{\footnotesize \mbox{longitudinal}}$) follows a pattern similar to that of a normal distribution. This observation is corroborated by the p-values, where the Shapiro-Wilk test does not reject the null hypothesis, with p-values $>0.05$, suggesting a possible normality in the dispersion indices obtained in the simulation study. This result indicates symmetry in the values and consistency in the pattern of the indices, regardless of the analyzed scenarios, by Figures \ref{IDML_HistM5} to \ref{IDML_HistM15}.

\begin{table}[H]
\centering
\footnotesize
\caption{Descriptive statistics related to the longitudinal dispersion index based on $1000$ sets of simulated data,  with groups composed of 10 individuals ($m=10$), for sample size $N=100$ }
\begin{tabular}{lrr|rr|rr}
\hline
 & \multicolumn{2}{c}{$j=3$} & \multicolumn{2}{c}{$j=4$} & \multicolumn{2}{c}{$j=5$ } \\ \cline{2-7} 
 & \multicolumn{6}{c}{$t=3$}  \\ \cline{2-7} 
 & $\sigma^{2} = 0.01$   & $\sigma^{2} =10$   & $\sigma^{2} = 0.01$   & $\sigma^{2} =10$   & $\sigma^{2} = 0.01$   & $\sigma^{2} =10$  \\ 
 \hline
Maximum   & 0.288 & 0.676  & 0.236 & 0.584 &  0.237  & 0.571 \\
Minimum   & 0.202 & 0.578  & 0.170 & 0.478 & 0.156  & 0.472 \\
Amplitude  & 0.086 & 0.097 & 0.065 & 0.105 & 0.081  & 0.098 \\
Mean  & 0.243 & 0.626 & 0.198 & 0.538 & 0.192  & 0.518 \\
Standard deviation & 0.013 & 0.015  & 0.010 & 0.015 & 0.010  & 0.015 \\
\hline
 & \multicolumn{6}{c}{$t=4$} \\ \cline{2-7} 
 & $\sigma^{2} = 0.01$   & $\sigma^{2} =10$   & $\sigma^{2} = 0.01$   & $\sigma^{2} =10$   & $\sigma^{2} = 0.01$   & $\sigma^{2} =10$  \\ 
 \hline
Maximum     & 0.279& 0.671  & 0.226 & 0.582 & 0.225 & 0.563 \\
Minimum   & 0.205 &  0.586 & 0.170 & 0.498 & 0.164  & 0.469 \\
Amplitude & 0.073 & 0.084  & 0.056 & 0.084 & 0.060  & 0.093 \\
Mean      & 0.243 & 0.626  & 0.198 & 0.538 & 0.192  & 0.516 \\
Standard deviation & 0.011 & 0.014  & 0.009 & 0.013 & 0.009  & 0.014 \\
\hline
\end{tabular}
\label{IDML_TabM10}
\end{table}

\begin{table}[H]
\centering
\footnotesize
\caption{Descriptive statistics related to the longitudinal dispersion index based on $1000$ sets of simulated data,  with groups composed of 15 individuals ($m=15$), for sample size $N=100$.}
\begin{tabular}{lrr|rr|rr}
\hline
 & \multicolumn{2}{c}{$j=3$} & \multicolumn{2}{c}{$j=4$} & \multicolumn{2}{c}{$j=5$ } \\  \cline{2-7} 
 & \multicolumn{6}{c}{$t=3$} \\ \cline{2-7}
 & $\sigma^{2} = 0.01$   & $\sigma^{2} =10$   & $\sigma^{2} = 0.01$   & $\sigma^{2} =10$   & $\sigma^{2} = 0.01$   & $\sigma^{2} =10$  \\ 
 \hline
Maximum   & 0.256 & 0.663  & 0.200 & 0.570 &  0.194  & 0.548 \\
Minimum   & 0.176 & 0.564  & 0.138 & 0.471 & 0.134  & 0.452 \\
Amplitude  & 0.079 & 0.099 & 0.061 & 0.098 & 0.059  & 0.095 \\
Mean   & 0.215 & 0.612 & 0.168 & 0.520  & 0.162 & 0.498 \\
Standard deviation & 0.012 & 0.015  & 0.010 & 0.014 & 0.010  & 0.014 \\
\hline
 & \multicolumn{6}{c}{$t=4$} \\ \cline{2-7} 
 & $\sigma^{2} = 0.01$   & $\sigma^{2} =10$   & $\sigma^{2}= 0.01$   & $\sigma^{2} =10$   & $\sigma^{2} = 0.01$   & $\sigma^{2}=10$  \\ 
 \hline
Maximum     & 0.249& 0.656  & 0.199 & 0.565 & 0.190 & 0.539 \\
Minimum   & 0.176 &  0.571 & 0.142 & 0.479 & 0.135  & 0.456 \\
Amplitude & 0.072 & 0.085 & 0.057 & 0.086 & 0.054  & 0.083\\
Mean     & 0.215& 0.612  & 0.169 & 0.520 & 0.162  & 0.498 \\
Standard deviation & 0.010 & 0.013 & 0.008 & 0.012 & 0.008  & 0.012 \\
\hline
\end{tabular}
\label{IDML_TabM15}
\end{table}

Considering the results obtained for scenarios with $m=10$ and $m=15$ individuals per group, presented in Tables \ref{IDML_TabM10} and \ref{IDML_TabM15} respectively, it was observed that, in both cases, the scenarios demonstrate degrees of equidispersion and overdispersion in a manner similar to that observed for $m=5$ individuals per group (Table \ref{IDML_TabM5}).

For the other scenarios, that is, with sample sizes of $N=200$ and $N=500$, results similar to those found for the sample size of $N=100$ were verified. This consistency in results across different sample sizes reinforces the reliability and robustness of the analyses conducted in the simulation study, providing reliable parameters for cases that exhibit degrees of equidispersion and overdispersion.

%For more details on the results of these scenarios, they are available at 

%\url{https://github.com/GabrielRPalma/PolytomousDataOverdispersionIndex}.

\begin{figure}[H]
\centering
\includegraphics[width=1.0\textwidth]{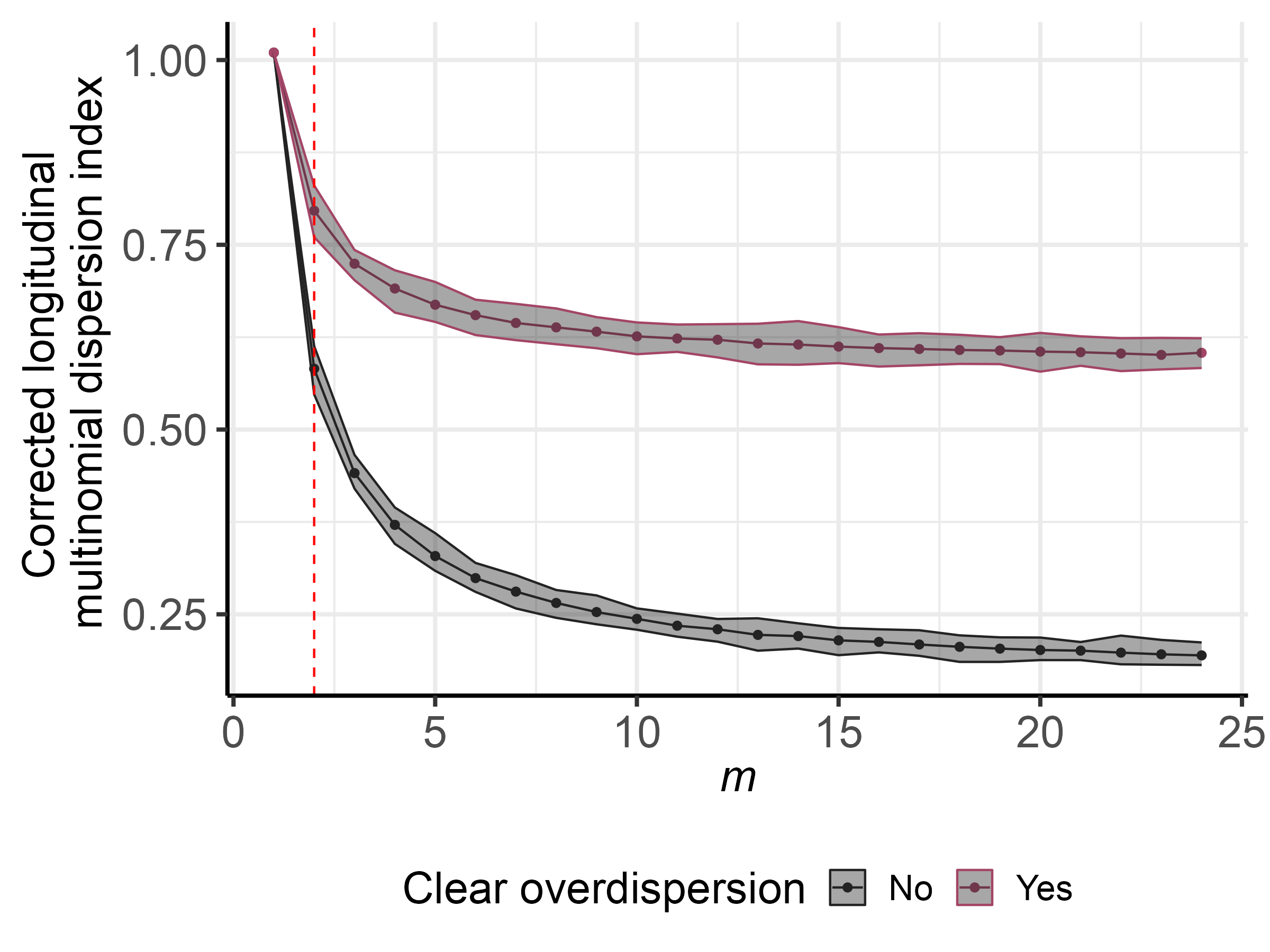}
 \caption{The average, $2.5\%$ and $97.5\%$ percentiles of the longitudinal multinomial dispersion index calculated for $J=3$, $t=10$ for different values of $m$. The dashed line indicates the case where $m = 2$.}
 \label{m_behaviour}
\end{figure}

%Finally, Figure~\ref{m_behaviour} highlights that the corrected longitudinal multinomial dispersion index was mainly designed for group scenarios with $m\geq 2$, and as $m$ increases the differences between the scenarios high degree of overdispersion ...

Finally, Figure \ref{m_behaviour} highlights that the corrected longitudinal multinomial dispersion index was mainly designed for group scenarios with $m=2$, as for $m=1$ it is not possible to differentiate cases of equidispersion and overdispersion. Furthermore, as $m$ increases, it becomes possible to distinguish scenarios that exhibit high or low degrees of overdispersion. That is, the closer the values of the index are to $1$, the more likely it is that a high degree of overdispersion occurs. On the other hand, values close to $0$ indicate a low degree of overdispersion in the data, regardless of the value of $m$ or the scenario analyzed.

\section{Case Study} \label{application}
An experiment developed by \cite{castro2016comportamento} from March to June $2014$, involving male pigs, was set up in a completely randomized design. This study aimed to evaluate the behaviour of these animals subjected to two different rearing conditions: with environmental enrichment and without environmental enrichment. In the enriched stalls, suspended chains and loose plastic containers were introduced, while the stalls without enrichment remained without objects.

This study used a group of male pigs in the growing phase, comprising approximately $90$ days, totalling $128$ animals, randomly distributed in $8$ pens, each with $16$ pigs ($m=16$). Data collection was carried out on days $1$, $8$, $14$, and $27$ of the study, always at half past eight in the morning, with the variable of interest being animal behaviour, categorized as ``resting'', ``eating'' and ``exploring''. The analyses are performed using the R software \citep{R}, through the \textit{mclogit} package \citep{elff2022package} for fitting the mixed generalized logits model.

\begin{figure}[H]
\centering
\includegraphics[width=0.8\textwidth]{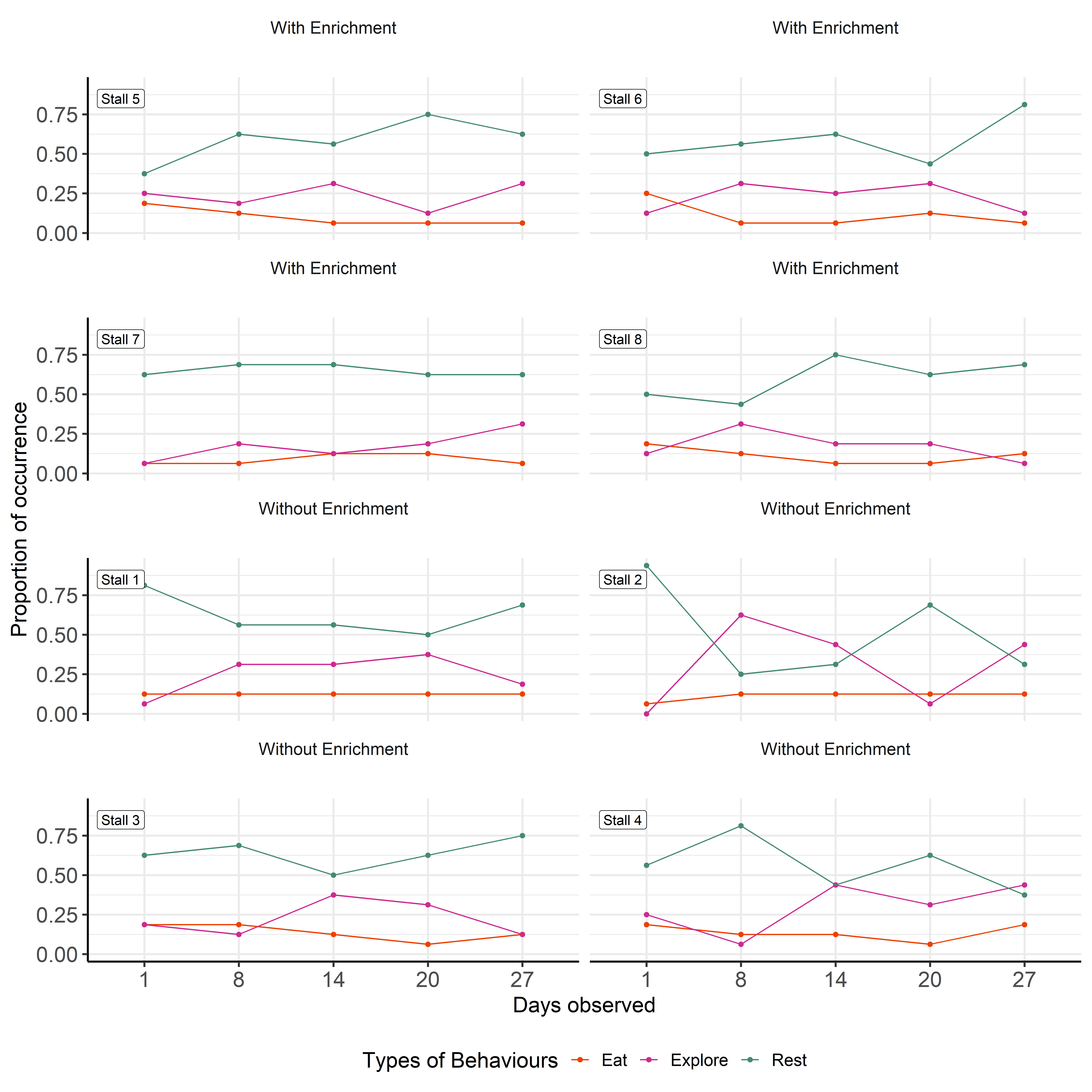}
 \caption{Occurrences of pig behaviour classifications over the observed days, considering the rearing conditions with and without environmental enrichment during the experiment conducted by \cite{castro2016comportamento} from March to June 2014.}
 \label{sec_apl1}
\end{figure}

Initially, an exploratory analysis was conducted to understand the behaviour of the data. In Figure \ref{sec_apl1}, there is evidence that, regardless of the environmental condition, the predominant behaviour was resting in most of the pens. It is also observed that there appears to be no difference in the occurrence of behaviours within the pens in the two treatment groups. In the SE environmental condition, there is evidently greater variability in the occurrence of behaviours within the pens. On the other hand, besides ``resting'', the behaviour ``exploring'' predominates in the CE environmental condition, indicating the interaction of the animals with objects in the stall.

Subsequently, mixed generalized logits models were fitted, adopting the canonical logistic link function. These models consider the factor of the type of environmental condition with two levels with environmental enrichment and without environmental enrichment, the day factor with $5$ levels, the effect associated with the interaction between the type of environmental condition and the day, and the random effects corresponding to the $8$ stalls, constituting the linear predictor. Thus, establishing the behaviour type ``resting'' as the reference category $(J=1)$, the complete model is defined by:

\begin{equation} \label{modcompapl}
\footnotesize{
\ln \left[\dfrac{\pi_{ijkt}}{\pi_{i1kt}}\right] = \eta_{ijkt}= \alpha_j + \beta_{jk}( \mbox{treat})_{jk} + \tau_{jt}(\mbox{day})_{jt} +(\beta \tau)_{jkt} (\mbox{treat} \times \mbox{day})_{jkt} + _i}
\end{equation}

\noindent where $j=2,3$ (eating and exploring), $k=1$ (without environmental enrichment), $t=1,2, \cdots, 5$, and $i=1, \cdots, 8$. In this context, $\alpha_j$ is the intercept for the $j$-th behaviour type category, $\beta_{jk} \mbox{treat}_{jk}$ is the effect associated with the $k$-th environmental condition, $\tau_{jt}\mbox{day}_{jt}$ is the effect associated with the $t$-th observed day, $(\beta \tau)_{jkt} (\mbox{treat} \times \mbox{day})_{jkt}$ is the effect associated with the interaction between the type of environmental condition and the observed day, and $u_i$ are the random effects associated with the $i$-th stall, where $u_i \sim N(0, \Sigma_j)$.

For the selection of covariates in the model, the nested model process was used, utilizing deviation analysis as a criterion, considering the following submodels:

\textbf{Model 1:} only with the random effect of the pen,

\begin{equation} \label{mod1}
\eta_{ijkt}= \alpha_j + _i.
\end{equation}

\textbf{Model 2:} with the addition of treatment effect to model (\ref{mod1}),

\begin{equation} \label{mod2}
\eta_{ijkt}= \alpha_j + \beta_{jk} (\mbox{treat})_{jk} + _i.
\end{equation}

\textbf{Model 3:} with the addition of the day effect to model (\ref{mod1})

\begin{equation} \label{mod3}
\eta_{ijkt}= \alpha_j + \tau_{jt}(\mbox{day})_{jt} + _i.
\end{equation}

\textbf{Model 4:} considering the treatment effect added in model (\ref{mod1}) and the day effect added to model (\ref{mod2}),

\begin{equation} \label{mod4}
\eta_{ijkt}= \alpha_j + \beta_{jk} (\mbox{treat}){jk} + \tau{jt}(\mbox{day})_{jt} + u_i.
\end{equation}

The sequential models described by equations (\ref{modcompapl}), (\ref{mod1}), (\ref{mod2}), (\ref{mod3}), and (\ref{mod4}) were fitted using the maximum likelihood method, and model selection was conducted using the nested model criterion. The degrees of freedom and the deviance are presented in Table \ref{modencaixados}.

\begin{table}[H]
\centering
\footnotesize
\caption{Selection of the linear predictor among the sequential models, through \textit{deviance} analysis, for the experimental data conducted by \citet{castro2016comportamento} from March to June 2014.}
\begin{tabular}{clccrr}
\hline
Model & Linear Predictor & Comparison & \begin{tabular}[c]{@{}c@{}}Degrees of\\ Freedom\end{tabular} & Deviance & p-value \\
\hline
1 & Intercept + Random Effect & & & & \\
2 & \begin{tabular}[c]{@{}l@{}}Environmental Condition \\ + Random Effect\end{tabular} & 1 $\times$ 2 & 8 & 12.51 & 0.12 \\
3 & Day + Random Effect & 2 $\times$ 3 & 6 & 9.08 & 0.16 \\
4 & \begin{tabular}[c]{@{}l@{}}Environmental Condition \\ + Day + Random Effect\end{tabular} & 3 $\times$ 4 & 8 & 12.78 & 0.11 \\
5 & \begin{tabular}[c]{@{}l@{}}Environmental Condition*Day \\ + Random Effect\end{tabular} & 4 $\times$ 5 & 11 & 8.89 & 0.84 \\ 
\hline
\end{tabular}
\label{modencaixados}
\end{table}

Based on Table \ref{modencaixados}, model 1 was selected, indicating that there is evidence that the environmental enrichment conditions and the days considered in the study do not interfere with the behaviour classification adopted by the pigs.

The estimated parameters and standard errors of the model are presented in Table \ref{Tab_est1}.

\begin{table}[H] 
\centering 
\caption{Estimated regression parameters of Model 1 for the experiment conducted by \citet{castro2016comportamento} from March to June 2014.} 
\begin{tabular}{lrr} 
\hline 
Parameter & Estimate & Standard Error \\
\hline 
$\alpha_2 \text{(intercept 2)}$ & -1.6361 & 0.1271 \\
$\alpha_3 \text{(intercept 3)}$ & -0.9162 & 0.0960 \\
\hline \end{tabular} \label{Tab_est1} \end{table}

\begin{table}[H]
\centering
\caption{Observed and expected variances after fitting the mixed generalized logits model including the rearing condition and the behaviour classification of pigs, based on the experiment conducted by \citet{castro2016comportamento} from March to June 2014.}
\begin{tabular}{clll|lll}
\hline
\multicolumn{1}{l}{}                                       & \multicolumn{3}{c|}{Observed Variance} & \multicolumn{3}{c}{Expected Variance} \\ 
\cline{2-7} 
\begin{tabular}[c]{@{}c@{}}Observed \\ Days\end{tabular} & Eating    & Exploring    & Resting    & Eating    & Exploring   & Resting   \\
\hline
1   & 1.14         & 2.12        & 8.41        & 2.11         & 1.84       & 3.78       \\
8   & 0.42         & 7.64        & 7.64        & 1.66         & 3.12       & 3.90       \\
14  & 0.27         & 3.27        & 4.68        & 1.46         & 3.39       & 3.95       \\
20  & 0.29         & 3.07        & 2.50        & 1.36         & 2.87       & 3.81       \\
27  & 0.50         & 5.43        & 7.93        & 1.56         & 3.00       & 3.81       \\ 
\hline
\end{tabular}
\label{sec_aplTab1}
\end{table}

Comparing the observed and expected variances after fitting the model of generalized mixed logits (Table \ref{sec_aplTab1}), it was noted that in some cases, the observed variances are close to the expected ones, while in others, the observed are lower than expected. Additionally, the value of the longitudinal multinomial dispersion index was $\mat{\Lambda}_{\text{longitudinal}} = 0.071$, this suggests a low degree of overdispersion.

On the other hand, given the variance analysis, in which in some cases the observed variances were lower than expected, there is evidence of a scenario with degrees of underdispersion. In light of these results, future work is proposed to develop an index that characterizes this phenomenon. Such an index would not only complement the current study but would also serve as a measurement and diagnostic tool in analyses of grouped nominal polytomous data where degrees of underdispersion of this phenomenon are evident. This index would assist in distinguishing between different levels of dispersion, facilitating the application of more appropriate statistical models for each scenario.

\section{Conclusion} \label{conclusion}

In this chapter, an initial exploration is proposed for constructing a longitudinal multinomial dispersion index for nominal polytomous data in a longitudinal context. The efficacy of this index was evaluated through a simulation study, which revealed differences between high and low equidispersion levels starting from $m=2$ individuals per group, highlighting that the index is valid only for grouped nominal polytomous data. Values close to zero indicate a low degree of overdispersion, while values near one indicate a high degree of overdispersion.

Moreover, it must be emphasized that diagnosing a high degree of overdispersion is essential to prevent erroneous conclusions. This fact underscores the necessity and importance of the proposed index. It should also be noted that in the presence of high overdispersion, marginal models are not suitable because they do not consider the extra variability. Caution is needed in transition models. Therefore, the approach of mixed generalized logit models is considered most appropriate as it addresses overdispersion and ensures the validity of statistical inferences.

The practical applicability of the index demonstrated in a study on the behaviour of pigs under different environmental enrichment conditions, not only confirmed its efficacy with real data but also emphasized its relevance for applied research, especially in agricultural and biological sciences.

As perspectives for future work, the development of the dispersion index as a diagnostic measure for individual data structures and the investigation of underdispersion, a less common phenomenon that deserves attention, particularly in agricultural sciences, are suggested.

\section{Acknowledgments}

This publication has emanated from research conducted with the financial support of the Coordination for the Improvement of Higher Education Personnel (CAPES), process number 88882.378344/2019-01, the National Council for Scientific and Technological Development (CNPq) and Science Foundation Ireland under Grant number 18/CRT/6049. The opinions, findings, and conclusions or recommendations expressed in this material are those of the authors and do not necessarily reflect the views of the funding agencies.

%\newpage
%\section*{Referências}
\addcontentsline{toc}{section}{Referências}
\bibliography{referencias/bibliografia}

\end{document}